\documentclass[aps,amsmath,amssymb,superscriptaddress,nofootinbib,onecolumn]{revtex4-2}
\usepackage{graphicx}
\usepackage{xcolor}
\usepackage[colorlinks,citecolor=blue,linkcolor=blue,urlcolor=blue,anchorcolor=blue]{hyperref}
\usepackage{longtable}
\usepackage{subfigure}
\usepackage{rotating}

\renewcommand\apj{Astrophys. J.}
\newcommand\apjl{Astrophys. J. Lett.}
\newcommand\apjs{Astrophys. J. Suppl.}
\newcommand\araa{Annu. Rev. Astron. Astrophys.}
\newcommand\aapr{Astron. Astrophys. Rev.}

\newcommand\aap{Astron. Astrophys.}

\newcommand\mnras{Mon. Not. R. Astron. Soc.}
\newcommand\physrep{Phys. Rep.}

\newcommand\apss{Astrophysics and Space Science}
\newcommand\etal{{\it et~al.}}
\newcommand{\ME}{the M\&E 2018 model}
  
\def\DM{{\rm DM}}
\begin{document}

\title{Distribution of fast radio burst dispersion measures in CHIME/FRB Catalog 1: implications on the origin of FRBs}

\author{Jianwei Zhang}
\affiliation{CAS Key Laboratory of FAST, National Astronomical Observatories, Chinese Academy of Sciences, Beijing 100101, China}
\affiliation{School of Physical Sciences, University of Chinese Academy of Sciences, Beijing 100049, China}

\author{Chengmin Zhang}
\email{zhangcm@bao.ac.cn}
\affiliation{CAS Key Laboratory of FAST, National Astronomical Observatories, Chinese Academy of Sciences, Beijing 100101, China}
\affiliation{School of Physical Sciences, University of Chinese Academy of Sciences, Beijing 100049, China}

\author{Di Li}
\email{dili@nao.cas.cn}
\affiliation{CAS Key Laboratory of FAST, National Astronomical Observatories, Chinese Academy of Sciences, Beijing 100101, China}
\affiliation{School of Physical Sciences, University of Chinese Academy of Sciences, Beijing 100049, China}
\affiliation{NAOC-UKZN Computational Astrophysics Centre, University of KwaZulu-Natal, Durban 4000, South Africa}
\affiliation{Research Center for Intelligent Computing Platforms, Zhejiang Laboratory, Hangzhou 311100, China}

\author{Wuming Yang}
\affiliation{Department of Astronomy, Beijing Normal University, Beijing 100875, China}

\author{Xianghan Cui}
\affiliation{CAS Key Laboratory of FAST, National Astronomical Observatories, Chinese Academy of Sciences, Beijing 100101, China}
\affiliation{School of Physical Sciences, University of Chinese Academy of Sciences, Beijing 100049, China}

\author{ChangQing Ye}
\affiliation{School of Physics and Astronomy, Sun Yat-sen University, Zhuhai 519082, China}

\author{Dehua Wang}
\affiliation{School of Physics and Electronic Science, Guizhou Normal University, Guiyang 550001, China}

\author{Yiyan Yang} 
\affiliation{School of Physics and Electronic Science, Guizhou Education University, Guiyang 550018, China}

\author{Shaolan Bi}
\affiliation{Department of Astronomy, Beijing Normal University, Beijing 100875, China}

\author{Xianfei Zhang}
\affiliation{Department of Astronomy, Beijing Normal University, Beijing 100875, China}

\date{\today}


\begin{abstract}

Recently, CHIME/FRB project published its first fast radio burst (FRB) catalog (hereafter, Catalog 1), which totally contains 536 unique bursts. 
With the help of the latest set of FRBs in this large-size catalog, we aim to investigate the dispersion measure (DM) or redshift ($z$) distribution of the FRB population, and solution of this problem could be used to clarify the question of FRB origin.
In this study, we adopted \ME \, to fit the observed $z$ distribution of FRBs in Catalog 1.
In \ME, we are mostly interested in the $\Phi(z)$ function, i.e., number of bursts per proper time per comoving volume, which is represented by the star formation rate (SFR) with a power-law index $n$.
Our estimated value of $n$ is $0.0_{-0.0}^{+0.6}$ ($0.0_{-0.0}^{+2.1}$) at the 68 (95) per cent confidence level, implying that the FRB population evolves with redshift consistent with, or faster than, the SFR.
Specially, the consistency of the $n$ values estimated by this study and the SFR provides a potential support for the hypothesis of FRBs originating from young magnetars.

\end{abstract}


\maketitle


\section{Introduction}\label{sec:intro}

Fast radio bursts (FRBs) have received much attention in astrophysics, since the discovery of the first FRB known as the Lorimer burst \cite{2007Sci...318..777L}.
FRBs are the millisecond-duration radio flashes that originate at cosmological distances in the most cases \cite[see Refs.][and references therein for pedagogical reviews]{2019ARA&A..57..417C,2019A&ARv..27....4P,2022A&ARv..30....2P}.
The origins of the most FRBs are still unknown, and a wide range of the theoretical progenitor models have been proposed to explain their properties \cite[see Ref.][and references therein]{2019PhR...821....1P}.
Remarkably, several recent observational breakthroughs provide support for a magnetar  \cite{1992ApJ...392L...9D,2008MNRAS.389L..66F,2011ApJ...734...44Z,2017ARA&A..55..261K,2018MNRAS.473.3204I} or a neutron-star (NS) origin, for at least some FRBs. 
For instance, the Canadian Hydrogen Intensity Mapping Experiment FRB Project (CHIME/FRB) \cite{2020Natur.587...54C} and STARE2 \cite{2020Natur.587...59B} simultaneously detected an extremely bright FRB-like signal (FRB 200428) from a known Galactic magnetar (SGR 1935+2154).
And, Ref.~\cite{2022Natur.607..256C} reported two FRBs (i.e., FRB 20210206A and FRB 20210213A) with sub-second periodicities between their multi-component pulses, as 2.8 and 10.7 ms, respectively, and these observed short periodicities suggested that the super-giant pulse from a NS is a possible explanation. 
And, Ref.~\cite{2021Natur.598..267L} reported the detection of 1652 independent bursts from FRB 121102, observed by the Five-hundred-meter Aperture Spherical radio Telescope (FAST) \cite{2006ScChG..49..129N,2018IMMag..19..112L}, this high emitting efficiency in the radio band suggested the mechanisms that invoke a NS magnetosphere.
Moreover, Ref.~\cite{2022Sci...375.1266F} investigated the polarization properties of five repeating FRBs and suggested that a complex environment (e.g., a supernova remnant or a pulsar wind nebula) may be close to the repeating FRBs.
Statistically, Refs.~\cite{2021RAA....21..211C,2022Ap&SS.367...66C} suggested that magnetars are favoured over gamma-ray bursts as FRB origins, by analyzing the FRB luminosity distributions. 
On the theoretical model side, FRBs as flaring magnetars have also been proposed \cite[e.g., Refs.][]{2014ApJ...797...70K,2016MNRAS.457..232C,2018ApJ...868L...4M,2018IJMPD..2744016P,2022ApJ...927....2K}.
In addition, if FRBs originate from young magnetars, they would be consistent with the star formation rate (SFR) \cite{2017ApJ...841...14M}, and several researches \cite[e.g., Refs.][]{2018MNRAS.480.4211M,2022MNRAS.510L..18J,2022MNRAS.509.4775J,2021MNRAS.501.5319A,2022MNRAS.512.2093A} in recent years have modelled the FRB population with respect to some power of the cosmic SFR.

Regardless of the origins and physical mechanisms of FRBs are uncertain now, FRBs have been proved to be an useful tool to probe the universe \cite{2016Natur.530..427L,2018NatAs...2..860L,2010HiA....15..131L}, such as probing properties of the intergalactic medium (IGM) \cite{2020Natur.581..391M}, constraining the electron density of our Galactic halo \cite{2020ApJ...895L..49P}, providing the constraint on the distributions of dark matter \cite{2020ApJ...900..122S}. 
And, the population properties of FRBs --- e.g., dispersion measure (DM), rotation measure (RM), sky distribution, pulse width, and luminosity, etc. --- are also themselves valuable to be investigated.
Among them, DM is of high interest as it contains the information about the distances of FRBs, thereby establishing the extragalactic origins for FRBs. 
The DM is defined as the integrated column density of free electrons $n_e$ along the line of sight, weighted by $(1+z)^{-1}$ for the cosmological sources, expressed as $\DM = \int_{0}^{l} n_e ds/(1+z)$, where $l$ is the path length, $z$ is the redshift \cite{2019ARA&A..57..417C,2019A&ARv..27....4P}.

Recently, Ref.~\cite{2021ApJS..257...59C} reported its first FRB catalog (hereafter, Catalog 1), including 536 bursts (62 bursts from 18 repeating sources and 474 one-off bursts), approximately as the quadruple of all previously published FRBs \footnote{\url{https://www.herta-experiment.org/frbstats/}}. 
Such large size of FRB samples, detected with a single telescope, is absolutely an opportunity for the entire population studies. 
For instance,
Ref.~\cite{2021ApJS..257...59C} compared the DM distribution for apparent non-repeaters and repeater in Catalog 1 by using Anderson-Darling (A-D) and
Kolmogorov-Smirnov (K-S) tests, and they found that it is consistent with being from the same underlying DM distribution for above two-type FRBs.
Ref.~\cite{2021ApJ...923....2J} investigated the FRB sky distribution in Catalog 1, and inferred that FRBs are isotropic cosmological sources.
And, Ref.~\cite{2021ApJ...922...42R} found that there is a FRB-galaxy correlation in range of $0.3 \leq z \leq 0.5$, regarding the FRBs in Catalog 1.

Benefiting from the publication of Catalog 1, in this paper, we aim to examine the DM (or redshift) distribution of the FRB population in Catalog 1, and our results are expected to provide insight into the origin of FRBs. 
This paper is organized as follows. 
In Section~\ref{sec:method}, our method of modelling is described. Results and discussions are given in Section~\ref{sec:result}.

\section{Method}\label{sec:method}

\subsection{The Selection of FRB Samples}\label{subsec:sample}

Visiting the data release website for Catalog 1 \footnote{\url{https://www.chime-frb.ca/catalog}}, there are in total 536 FRBs that have been published \cite[refer to Ref.][]{2021ApJS..257...59C}.
We described the criteria and results of selecting our FRB samples from Catalog 1 in the followings:

\begin{enumerate}

\item Given that apparently non-repeating sources and repeating ones may have different origins \cite{2021MNRAS.500.3275C,2020RNAAS...4...98P}, then, in order to maintain the homology property of our samples, only 474 FRBs (with \textsf{repeater\_name=-9999}) \cite{2021ApJS..257...59C} from so-far non-repeaters were selected.

\item We excluded 36 FRBs (with \textsf{excluded\_flag=1}) \cite{2021ApJS..257...59C} from previous samples, leaving 438 FRBs. This is because these events were detected during the stages of pre-commissioning or software upgrades, which may lead to a low-sensitivity.

\item We also excluded 3 FRBs (i.e., FRB20190210D, FRB20190125B, FRB20190202B) \cite{2021ApJS..257...59C} that were detected in the far side-lobes of CHIME, since this regime is poorly understood.
\end{enumerate}

Overall, after above selection process, there are totally 435 FRBs included in our samples.

\subsection{DM Distribution Model} \label{subsec:model}

\begin{figure}
\centering\includegraphics[width=0.8\textwidth]{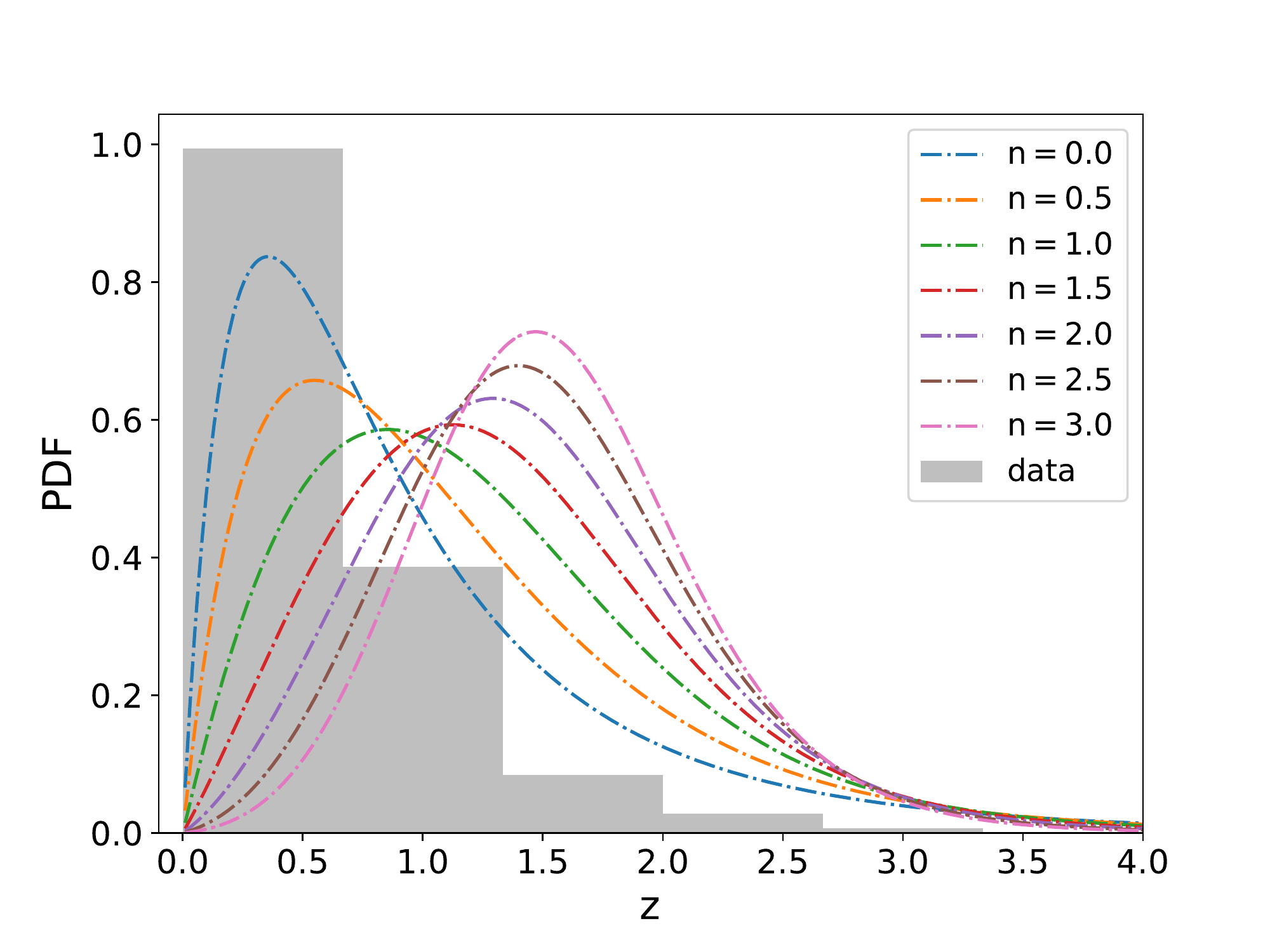}
\caption{Comparisons of the modelled redshift $z$ distributions with a set of representative $n$ values (i.e., $n \in [0.0,3.0]$, $\Delta n = 0.5$) and the observed $z$ histogram for FRBs in Catalog 1, in which the number of bins is 6 from $z=0$ to $z=4$.}
\label{fig1}
\end{figure}

Generally, the observed $\DM$ for a given FRB can be written as the summation of four individual components \cite{2019ARA&A..57..417C,2019A&ARv..27....4P}:
\begin{equation}
\label{eq1}
\DM_{\rm Obs} (z) = \DM_{\rm MW, ISM} + \DM_{\rm MW, Halo} + \DM_{\rm IGM} (z) + \frac{\DM_{\rm Host}}{1+z} \, ,
\end{equation}
where $\DM_{\rm MW, ISM}$ is the contribution from the interstellar medium (ISM) of our galaxy (i.e., Milky Way), usually estimated by the electron density models, e.g., YMW16 \cite{2017ApJ...835...29Y} or NE2001 \cite{2002astro.ph..7156C}. In our subsequent analysing, the values from YMW16 were adopted, and it will not affect our conclusion if we used NE2001; 
$\DM_{\rm MW, Halo}$ and $\DM_{\rm Host}$ are the contribution from our Galactic halo and the host galaxy, respectively. They are assumed to be 50 pc cm$^{-3}$ in our model, similar to the treatment of Ref.~\cite{2020Natur.581..391M}.
$\DM_{\rm IGM} (z)$ is the contribution from the intergalactic medium (IGM), which can be roughly calculated by the well-known $\DM-z$ relation (i.e., the Macquart relation) \cite{2020Natur.581..391M}, as $\DM_{\rm IGM} (z) \approx 900 z$ .
Given above assumptions, Equation~(\ref{eq1}) remains only one variable, i.e., the redshift $z$. Thus, one can calculated the $z$ value for each FRB, then obtained the $z$ distribution for the FRB samples in Catalog 1 (see FIG.~\ref{fig1}), which can be regarded as the observations to some extent.

The $z$ distribution of FRBs is related to the properties of FRB luminosity distributions, their evolving comoving density $\Phi(z)$ and the properties of the radio telescope, especially its sensitivity. 
The interaction between these functions will provide the observed $z$ distribution.
Noted that, since Ref.~\cite{2021ApJS..257...59C} pointed that the selection effects in DM are modest for CHIME/FRB, these effects are not taken into account in this study.
Here, we adopted the model introduced by Ref.~\cite{2018MNRAS.480.4211M} (hereafter, \ME) to fit the observed $z$ distribution.
In \ME, the redshift distribution in a survey of limiting fluence, $F_{0}$, is given by,
\begin{equation}
    \begin{split}
    \dfrac{dR_{F}}{dz}&(F_{\nu} > F_{0},z) = 4 \pi D_H^5\left(\dfrac{D_M}{D_H}\right)^4 \dfrac{(1+z)^{\alpha-1}}{E(z)} \Phi(z) \\
    & \times \dfrac{(1+z)^{2-\alpha}}{4 \pi D_{L}^{2}(z)}
    \begin{cases}
        0 & F_{0} > F_{\mathrm{max}} \\
        \left(\dfrac{F_{\mathrm{max}}^{1-\gamma} - F_0^{1-\gamma}}{F_{\mathrm{max}}^{1-\gamma} - F_{\mathrm{min}}^{1-\gamma}}\right) & F_{\mathrm{min}} \le F_{0} \le F_{\mathrm{max}} \, ,\\
        1 & F_{0} < F_{\mathrm{min}}
    \end{cases}
    \end{split}
    \label{eq:RedshiftDistribution}
\end{equation}
where the minimum, $F_{\mathrm{min}}$, and maximum, $F_{\mathrm{max}}$, fluences are related to the corresponding minimum, $E_{\mathrm{min}}$, and maximum, $E_{\mathrm{max}}$, source energies, respectively, via  $E_{[\text{min/max}]} = 4 \pi D_{L}^{2}(z) F_{[\text{min/max}]}/((1+z)^{2-\alpha})$, and in which, the factor of $(1+z)^{-\alpha}$ accounts for the K-correction \cite{2002astro.ph.10394H} due to the fact that the source emitted its radiation in a different band from that observed by the telescope. 
And, the definitions of other associated symbols were listed in Table~\ref{tab:ListOfSymbols}.

Here, the form of the $\Phi(z)$ function in Equation~(\ref{eq:RedshiftDistribution}), i.e., number of bursts per proper time per comoving volume, was adopted from Refs.~\cite{2022MNRAS.510L..18J,2022MNRAS.509.4775J} with a power-law index $n$,
\begin{eqnarray}
\Phi(z) & = & \frac{\Phi_0}{1+z} \left( \frac{{\rm SFR}(z)}{{\rm SFR}(0)} \right)^n \, ,
\label{eq:phiz}
\end{eqnarray}
where SFR$(z)$ was taken from Ref.~\cite{2014ARA&A..52..415M},
\begin{eqnarray}
{\rm SFR}(z) & = & 1.0025738 (1+z)^{2.7} \left(1 + \left(\frac{1+z}{2.9}\right)^{5.6}\right)^{-1} \, .
\label{eq:sfr_n}
\end{eqnarray}

In our case, regarding the parameters in Equations~(\ref{eq:RedshiftDistribution}-\ref{eq:sfr_n}) , $F_{0} \approx 5$ Jy ms for CHIME/FRB \cite{2021ApJS..257...59C}; 
for the sake of simplicity, $E_{\mathrm{min}}$ and $E_{\mathrm{max}}$ are fixed to their general values from Refs.~\cite{2022MNRAS.510L..18J,2022MNRAS.509.4775J}, i.e., as $10^{30.0}$ erg and $10^{41.7}$ erg, respectively;
the fluence spectral index $\alpha$, defined as $F_{\nu} \propto \nu^{-\alpha}$, was found to be $\alpha=1.4 \pm 0.2$ for FRBs in Catalog 1 by Ref.~\cite{2021ApJS..257...59C}.
This $\alpha$ value has been consistent with those of Refs.~\cite{2022MNRAS.510L..18J,2022MNRAS.509.4775J,2021MNRAS.501.5319A,2022MNRAS.512.2093A}, as $\alpha \approx 1.5$. 
Besides, Refs.~\cite{2022MNRAS.510L..18J,2022MNRAS.509.4775J} showed that $\alpha$ is strongly correlated with the index $\gamma$. 
Therefore, we fixed at $\alpha=1.5$ and $\gamma=1.5$ in this study, as the same values from the best-fit model of Ref.~\cite{2021MNRAS.501.5319A}.

Overall, after above reductions, our model remains only one free parameter, i.e., the index $n$. Here, we examined the models in the range of $n \in [0.0,3.0]$, with the interval $\Delta n = 0.1$, to represent the observed $z$ distribution for FRBs in Catalog 1.

\section{Results and Discussions}\label{sec:result}

\begin{figure}
\centering\includegraphics[width=0.8\textwidth]{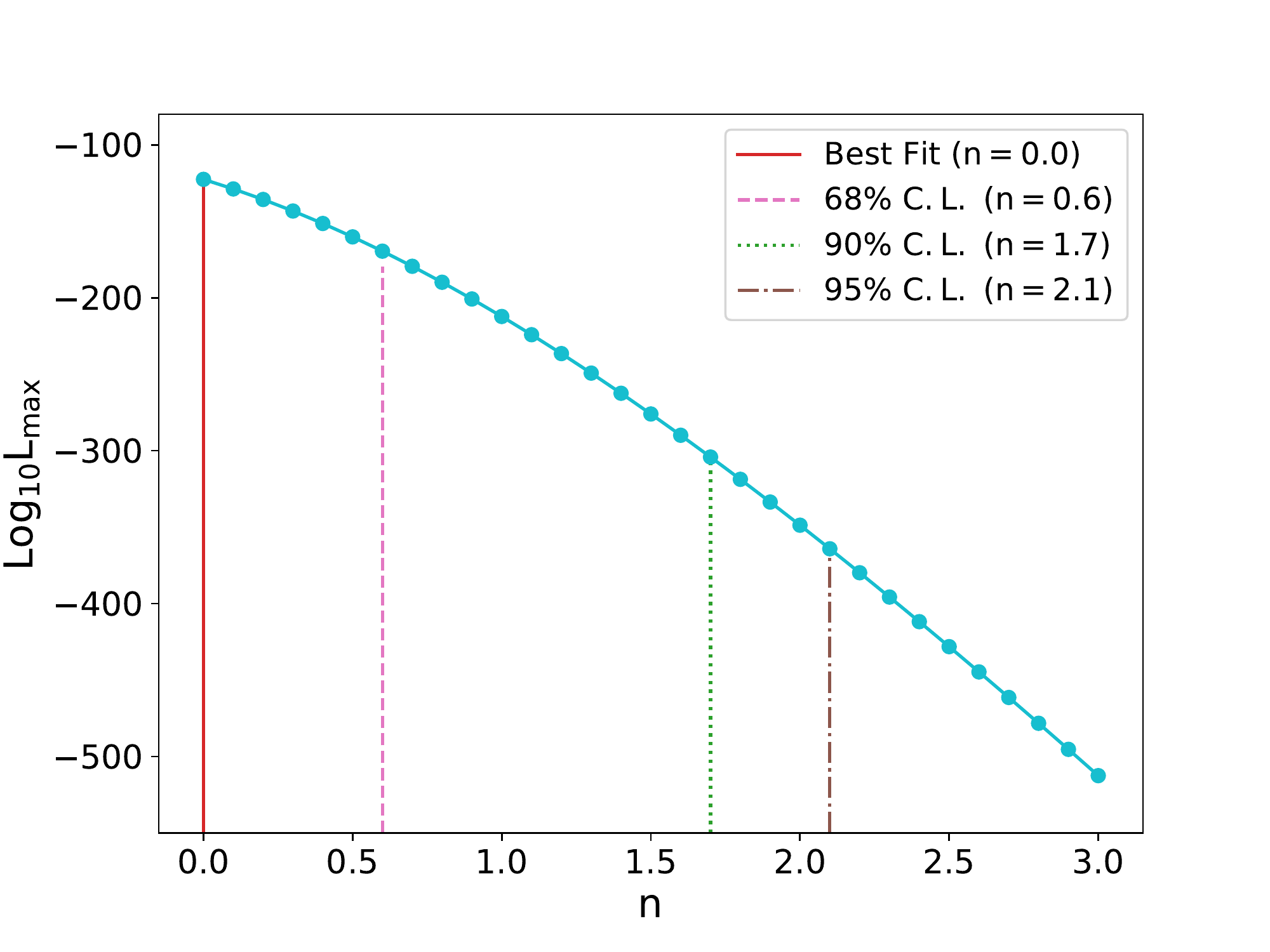}
\caption{Maximum likelihoods as a function for the parameter $n \in [0.0,3.0]$, $\Delta n = 0.1$, while the best-fitting value and confidence limits are indicated by the vertical lines calculated using Wilks' theorem \cite{Wilks1962}.}
\label{fig2}
\end{figure}

As described in Section~\ref{sec:method}, we modelled the $z$ distributions over the range of $0.0 \leq n \leq 3.0$, $\Delta n = 0.1$, and the results of modelling with the observed histogram were shown in FIG.~\ref{fig1}. 
In this study, the estimation of the parameter $n$ (i.e., the $\Phi(z)$ function) is of most interest.
Here, we calculated the best-fitting value and confidence limits of $n$ by using the maximum-likelihood (ML) method \cite{Wilks1962}, similar to that of Ref.~\cite{2022MNRAS.510L..18J}.
As shown in FIG.~\ref{fig2}, our estimated value is $n=0.0_{-0.0}^{+0.6}$ ($0.0_{-0.0}^{+2.1}$) at the 68 (95) per cent confidence level (C.L.).

Our estimates on the parameter $n$ suggest that the FRB population evolves with redshift consistent with, or faster than, the SFR.
And, our estimates of $n$ are roughly compatible with those of Refs.~\cite{2022MNRAS.510L..18J,2022MNRAS.509.4775J,2021MNRAS.501.5319A,2022MNRAS.512.2093A}. 
For instance, Ref.~\cite{2022MNRAS.510L..18J} found $n=1.67_{-0.40}^{+0.25}$ or $0.73_{-0.30}^{+0.30}$ at the 68 per cent C.L. depending on the interpretation of the spectral properties of the FRB, when they investigated the FRBs detected by the ASKAP and Parkes radio telescopes.
In particular, the consistency of the $n$ estimation of this study and the SFR would also support the hypothesis that FRBs may originate from young magnetars, for more information about the theoretical FRB-magnetar models, refer to Ref.~\cite{2019PhR...821....1P}. 
We are planning to further model the FRB redshift (or dispersion measure) distributions in the future, by jointly considering FRBs detected by a series of radio telescopes, which is expected to help us clarify the question of FRB origin more accurately.

%
\begin{acknowledgments}

This research is supported by the National Natural Science Foundation of China NSFC (11988101, 11773005, U2031203, U1631236, 11703001, U1731238, U1938117, 12163001, 11725313, 11721303), the International Partnership Program of Chinese Academy of Sciences grant No. 114A11KYSB20160008, the National Key R\&D Program of China No. 2016YFA0400702, and the Subsidy project of the National Natural Science Foundation (Grant No. 2021GZJ006). J.W.Z. acknowledges support from Jian-Wei Mountain House. We thank the anonymous referee for the critical comments and suggestions that have significantly improved the quality of the paper.

\end{acknowledgments}




%

\clearpage

\appendix*

\section{Model Symbol Definitions}\label{sec:append}

\begin{table*}[h]
\caption{For convenience purpose, the symbol definitions of \ME \cite{2018MNRAS.480.4211M} were provided, reproduced from Refs.~\cite{2021MNRAS.501.5319A,2022MNRAS.512.2093A}.}
\begin{tabular}{ll}
    \hline
    \hline
    Symbol                      & Definition \\
    \hline
    ${\rm DM}$                  & Dispersion measure \\
    $z$                         & Redshift \\
    $c$                         & Speed of light in vacuo \\
    $H_{0}$                     & Hubble constant at the present epoch \\
    $E(z)$                      & Dimensionless Hubble parameter : $E(z)=\sqrt{\Omega_{m} (1+z)^{3} + \Omega_{k} (1+z)^{2} + \Omega_{\Lambda}}$ \\
    $H(z)$                      & Hubble constant at an arbitrary redshift $z : H(z) =H_{0} E(z)$ \\
    $D_{H}$                     & Hubble distance : $D_{H}=c/H_{0}$ \\
    $D_{M}$                     & Comoving distance : $D_{M}=D_{H}\int_{0}^{z}\left(1/E(z)\right)dz$ \\
    $D_{L}$                     & Luminosity distance : $D_{L}=(1+z)D_{M}$ \\
    $R_{F}$                     & Total (fluence) differential FRB event rate in the observer's frame \\
    $\Omega_m$                  & Matter density (baryonic and dark) \\
    $\Omega_{\Lambda}$          & Vacuum density \\
    $\Omega_k$                  & Spatial curvature density \\
    $\Omega_b$                  & Baryonic matter density \\
    $\alpha$                    & Fluence spectral index defined such that $F_{\nu} \propto \nu^{-\alpha}$ \\
    $\gamma$                    & Energy power-law index \\
    $F_{0}$                     & Fluence survey limit at $\mathrm{DM}=0$ \\
    $F_{\nu}$                   & Fluence (energy spectral density per unit area) \\
    $F_{\mathrm{min}}$          & Minimum fluence for luminosity function \\
    $F_{\mathrm{max}}$          & Maximum fluence for luminosity function \\
    $E_{\nu}$                   & Spectral energy density \\
    $E_{\mathrm{min}}$          & Lower spectral energy density bound for the event rate energy function \\
    $E_{\mathrm{max}}$          & Upper spectral energy density bound for the event rate energy function \\
    $\displaystyle dR_{F}/dz$   & Fluence-based redshift distribution \\
    $\Phi(z)$                   & Bursts per proper time per comoving volume \\
    $n$                         & The power-law index of $\Phi(z)$ function \\
    ${\rm SFR}(z)$                   & Star formation rate \\
    \hline
\end{tabular}
\label{tab:ListOfSymbols}
\flushleft
\textbf{Notes.} In this study, we adopt a $\Lambda$CDM universe with cosmological parameters from Ref.~\cite{2014A&A...571A..16P}, i.e., $(\: H_{0},\: \Omega_b,\: \Omega_m,\: \Omega_{\Lambda},\: \Omega_k) = (70 \: \mathrm{km \: s}^{-1} \mathrm{Mpc}^{-1},\: 0.049,\: 0.318,\: 0.682,\: 0)$, similar to that of Refs.~\cite{2021MNRAS.501.5319A,2022MNRAS.512.2093A}.
\end{table*}

\end{document}